# Taylor & Francis Word Template for journal articles

Balwin Bokor[a]*, Klaus Altendorfer[a], Andrea Matta[b]

[a]Department for Production and Operation Management, University of Applied Science Upper Austria, A-4400 Steyr, Austria

[b]Department for Mechanical Engineering, University Politecnico di Milano, I-20133 Milano, Italy

*corresponding author: E-Mail: balwin.bokor@fh-steyr.at / Phone: +43 6644465335

**BALWIN BOKOR** is Research Associate in the field of Operations Management at the University of Applied Sciences Upper Austria. He is a PhD candidate and his research interests are, production planning, discrete event simulation, scheduling and energy simulation. His email address is balwin.bokor@fh-steyr.at

**KLAUS ALTENDORFER** is Professor in the field of Operations Management at the University of Applied Sciences Upper Austria. He received his PhD degree in Logistics and Operations Management and has research experience in simulation of production systems, stochastic inventory models and production planning. His e-mail address is klaus.altendorfer@fh-steyr.at

is Professor in the Department for Mechanical Engineering at the University Politecnico di Milano. His areas of expertise include Industrial Engineering with a focus on planning and scheduling, utilizing optimization techniques, heuristics, and simulation. His e-mail address is: andrea.matta@polimi.it.

# Optimizing Energy Consumption in Stochastic Production Systems: Using a Simulation-Based Approach for Stopping Policy

In response to the escalating need for sustainable manufacturing, this study introduces a *Simulation-Based Approach (SBA)* to model a stopping policy for energy-intensive stochastic production systems, developed and tested in a real-world industrial context. The case company – an energy-intensive lead-acid battery manufacturer – faces significant process uncertainty in its heat-treatment operations, making static planning inefficient. To evaluate a potential sensor-based solution, the *SBA* leverages simulated sensor data (using a Markovian model) to iteratively refine Bayesian energy estimates and dynamically adjust batch-specific processing times. A full-factorial numerical simulation, mirroring the company's 2024 heat-treatment process, evaluates the *SBA's* energy reduction potential, configuration robustness, and sensitivity to process uncertainty and sensor distortion. Results are benchmarked against three planning scenarios: (1) *Optimized Planned Processing Times (OPT)*; (2) the company's *Current Baseline Practice*; and (3) an *Ideal Scenario* with perfectly known energy requirements. *SBA* significantly outperforms *OPT* across all tested environments and in some cases even performs statistically equivalent to an *Ideal Scenario*. Compared to the *Current Baseline Practice*, energy input is reduced by 14–25%, depending on uncertainty and sensor accuracy. A Pareto analysis further highlights *SBA's* ability to balance energy and inspection-labour costs, offering actionable insights for industrial decision-makers.

# 1. Introduction

The surge in global energy consumption, driven by population growth and rising living standards, highlights an urgent need for sustainable manufacturing practices. In 2021, the industrial sector accounted for 26% and 30% of final energy consumption in the European Union and the USA, respectively (Eurostat 2023; U.S. Environmental Protection Agency 2023). Moreover, Data from the U.S. Energy Information Administration (2022) indicates that in 2021, using one kilowatt-hour of electricity in the USA released around 400g of carbon dioxide. As the world attempts to meet the ambitious 1.5-degree Celsius climate goal outlined in the Paris Agreement, reducing energy consumption and associated emissions in manufacturing becomes imperative, especially for energy-intensive products.

*1.1 Energy Reduction in Production Systems*

To address this challenge, various energy- efficiency strategies have been explored, particularly in energy-intensive manufacturing environments. A promising approach for improving energy efficiency implements policies that eliminate unnecessary energy consumption. Manufacturing processes often include energy-intensive operations, such as operating ovens, heat-treatment chambers, and drying systems, which might be repeated until a product achieves a specific technological feature. The challenge in these processes lies in accurately detecting when the target feature has been achieved, allowing the process to be terminated at the optimal time. Detecting it too early may result in a product that does not meet the required specifications, necessitating rework, while too late leads to unnecessary energy consumption. Thus, precise timing to balance product quality and energy efficiency is necessary. This challenge, referred to as the stopping problem in optimal control literature, is typically stochastic, involving variability in the measurement of the product feature and the manufacturing process execution (Aries and Shiryaev 2007). A *stopping policy* aims to determine the optimal time to halt a process, considering its current state and conditions to minimize costs or maximize rewards (Oh and Özer 2016). By identifying the moment when additional energy input no longer contributes to desired characteristics, such policies prevent waste and support sustainable manufacturing. Many stopping policies focus on controlling machine states, such as switching machines on/off or into standby. For instance, Weinert and Mose (2014), demonstrated through simulations that advanced standby strategies could achieve energy savings of up to 53% compared to conventional usage. Similarly, Frigerio and Matta

(2013) included warm-up periods in machine state policies, achieving energy reductions of up to 53% in real-world applications. However, these strategies often ignore product quality requirements and focus purely on operational efficiency, such as production rate, machine utilization, and costs.

In research, most analytical stopping policies assume deterministic conditions as in Cui et al. (2021) or Sun et al. (2020), which does not mirror real-world production environments. Altendorfer et al. (2016) highlight that actual production systems operate under uncertainty. In such uncertain environments, estimating stochastic variables based on measured data – such as the minimum required energy to achieve desired outcomes – becomes crucial for decision-making. Advances in Internet of Things and sensor technologies have significantly improved data collection (Wang et al. 2019), making real-time analysis feasible. Thereby, sensors play a key role in capturing variability and supporting better decision-making in stochastic settings.

*1.2 Real-World Case*

To illustrate this, we examine an industrial case where such a stopping problem arises: a heat-treatment process in lead-acid battery production. This energy-intensive process, used by a European lead-acid battery manufacturer, is subject to uncertainties such as varying moisture content, lead oxide paste thickness, and thermal losses within the heat-treatment chamber. These affect the minimum required energy to complete chemical curing and achieve target humidity – making it difficult to set deterministic planned processing times that are optimal for every batch. When times are too short, rework is required. When they are too long, energy is wasted. While the problem can be generalized as a stopping policy challenge, it cannot be solved analytically due to the complex stochastic nature and interdependence of process variables. Currently, the company applies static, experience-based planned processing times, favoring longer processing to mitigate the risk of defects. While this reduces rework, it leads to unnecessary energy consumption. The potential for batch-specific processing times, dynamically adapted using real-time sensor data, offers a promising way to reduce energy input without compromising product quality. To explore this potential, this article proposes a *Simulation-Based Approach (SBA)* that iteratively adjusts batch-specific processing times based on simulated sensor data. A Markovian model captures sensor behavior and generates time-series data for Bayesian estimation of the minimum required energy.

These estimates define an evolving energy threshold, which is then used to adjust the expected batch-specific remaining processing time.

*1.3 Model Contribution*

While the approach is particularly geared towards the heat-treatment processes of the company, it is designed to be generic and adaptable to other production systems operating under uncertain environments. Thus, the proposed approach offers both scientific and managerial value.

From a scientific perspective this work contributes: (1) A novel stopping policy for stochastic production systems utilizing an *SBA* to iteratively adjust processing times. (2) A generic uncertainty estimation approach combining a Markovian sensor-behavior model with Bayesian statistics used for energy estimation. From a managerial viewpoint, this work provides: (A) A practically tested approach for minimizing energy consumption in stochastic production systems, demonstrated in a real-world environment and benchmarked against three planning scenarios: *Optimized Planned Processing Times (OPT)*, where fixed planned processing times are applied uniformly; *Current Baseline Practice*; and an *Ideal Scenario*, with perfectly known energy requirements. (B) Key insights into the roles of sensor accuracy and process stability. (C) Decision-support information highlighting the trade-off between energy and personnel costs for rework inspections, visualized through a Pareto analysis.

This leads to the following research questions:

1. How can a stopping policy use simulated sensor data and Bayesian statistics to iteratively refine energy estimates and adjust remaining processing times?
2. What is the energy reduction potential of batch-specific processing times compared to the *Current Baseline Practice* and *OPT*, and how does it perform compared to an *Ideal Scenario* with perfectly known energy requirements?
3. How do varying levels of process uncertainty and sensor distortions impact the effectiveness of the *SBA* in reducing energy consumption and the respective parameters?
4. How is the trade-off between energy costs and personnel costs for rework affected by the *SBA* in comparison to *OPT*?

The paper is structured as follows: *Section 2* reviews relevant literature on energy-oriented production planning and scheduling through various approaches. *Section 3* outlines the case study and formal problem. *Section 4* introduces the *SBA*, including state modeling and an illustrative run. *Section 5* describes the simulation model based on which *Section 6* covers the numerical study. *Section 7* discusses results and benchmarking. *Section 8* concludes with key findings and future directions.

## 2. Related Literature

Production planning and scheduling significantly influences energy consumption in manufacturing. Terbrack et al. (2021) highlighted load management and supply orientation as central themes in energy-oriented production research. Biel and Glock (2016) detailed decision support models for lot-sizing and job allocation, whereas Fernandes et al. (2022) and Wenzel et al. (2018) reviewed energy-efficient job-shop scheduling and energy simulation methods, respectively. A synthesis of these comprehensive overviews reveals that energy reduction can be achieved at multiple levels in manufacturing. Reich-Weiser et al. (2010) identified three key levels for energy reduction in manufacturing: the device level, the broader enterprise level, and supply chain activities. Since production planning and scheduling decisions primarily impact the device level, Duflou et al. (2011) pinpointed three key strategies at this level: optimized machine tool design, improved process control (including changing the machine's state), and process/machine tool selection (including scheduling decisions). Across all three strategies, energy reduction can be achieved through different approaches: Rule-based decision-making, Mathematical optimization, Simulation-based approaches, Advanced analytics, or combinations of them to maximize performance.

*Rule-based decision-making* offers a simple way to enhance energy efficiency by utilizing domain knowledge without the need to model the full optimization problem. Mouzon et al. (2007) introduced dispatching rules to deactivate idle non-bottleneck machines. Bokor et al. (2024) developed dispatching rules balancing energy costs and tardiness, and Frigerio and Matta (2014) implemented threshold-based machine state controls to reduce idle energy consumption. Choi (2016) leveraged real-time information for a single-machine system with sequence-dependent setups to create energy-efficient dispatching rules. However, the dynamic nature of production systems often calls for more sophisticated algorithmic approaches. Genetic programming techniques have been applied successfully to generate rules from historical scheduling data for a Capacitated Lot-Sizing Problem by Hein et al. (2018) and by Braune et al. (2022) for a flexible job shop production system. Although *rule-based decision-making* contribute to energy efficiency and can deliver rapid solutions, they are not guaranteed to yield fully optimal results, particularly in complex production system environments.

*Mathematical optimization* techniques provide precise solutions for energy-efficient production planning and scheduling. Dai et al. (2019) reduced lead time and energy consumption through mixed-integer programming, while Wang et al. (2020) applied an epsilon-constraint method that incorporated time-based tariffs and on-site renewable energy integration. Energy uncertainty was addressed through stochastic programming by Golari et al. (2017) using scenario trees and solved the model via a Benders-type algorithm. Wichmann et al. (2019a, 2019b) extended the general lot-sizing and scheduling problem by incorporating energy tariffs, machine states, and multiple energy sources to minimize overall production costs. While, Saberi-Aliabad et al. (2020) further advanced the field of time-based tariffs by developing a mathematical model for scheduling jobs on unrelated parallel machines to minimizing energy costs. Although *mathematical optimization* can yield optimal solutions and provide valuable lower and upper bounds if halted early, it is often computationally demanding (Glover et al. 1999). This encourages the use of heuristics as exemplified by Vallejos-Cifuentes et al. (2019), who utilized a genetic algorithm to solve an energy-related optimization problem.

*Simulation-based approaches* effectively address the stochastic variability and uncertainty in production systems, as demonstrated by Stich et al. (2014) and Mousavi et al. (2016), who integrated energy prices and consumption models into simulation-based frameworks. Bokor and Altendorfer (2023b) specifically optimized planned processing times using simulation and energy input curves. Moreover, simulation serves a valuable test environment to compare the performance of different developed approaches – including algorithms, optimization problems, and heuristics – as shown by Felberbauer et al. (2012) or Seiringer et al. (2024). Combining simulation with optimization allows for capturing stochastic system behaviour while leveraging mathematical optimization (Glover et al. 1999). Heinzl et al. (2013) present frameworks for energy-efficient production via simulation-based optimization and data synchronization. Similarly, Sobottka et al. (2020) applied a simulation-based multi-criteria optimization to schedule heat-treatment processes in metal casting, considering energy costs and order batching under complex constraints.

*Advanced analytics* encompasses a broad range of techniques – such as machine learning, Markovian models, and Bayesian statistics – and offers new approaches to tackle uncertainty in production systems. Yamashiro and Nonaka (2021) used machine learning models to estimate complex processing time distributions for improved scheduling. Wei and Wang (2016) modelled a single-machine system via a Markov Decision Process incorporating machine states and surplus-dependent transitions to reduce energy consumption. Likewise, Li et al. (2019) applied a buffer-based Markovian model to estimate potential throughput loss and a genetic-based control algorithm to balance throughput and energy use. Fernlund (2010) used Bayesian statistics to minimize scrap by developing probability distributions based on prior and prototype data.

A pivotal element for the success of these approaches, particularly advanced analytics and data-driven methods, is high-quality information. According to Busert and Fay (2018) granularity, actuality, and accuracy are vital dimensions of information quality essential for reliable and effective planning. Sensors are primary instruments for capturing high-quality data, enhancing stochastic variable estimation. For instance, Ji et al. (2007) modeled sensor-driven state sequences as a Markov Decision Process to iteratively update and improve data reliability. Recent studies, such as Bokor and Altendorfer (2023a) highlighted the potential of sensor data integration into simulations to optimize batch-specific processing times, though challenges remain in achieving effective implementation across varying energy requirement ranges.

## 3. Problem Description

This section begins with an introduction to the heat-treatment process, followed by a description of the necessitated process steps and the corresponding energy input curves related to the case company. Subsequently, we present an abstracted formulation of the energy optimization problem faced by the company.

### 3.1 Heat-Treatment Process

The company studied produces lead-acid batteries for the automotive and industry aftermarket. In its heat-treatment process, lead plates – major component of lead-acid batteries – are produced. Once production orders are released, lead plates are casted and pasted with lead oxide, batched onto pallets, and buffered before facing the heat-treatment. Depending on the plate type, one of four heat-treatment programs is applied to ensure proper chemical curing and achieve the desired humidity level. Chemical curing strengthens the lead oxide paste's microstructure to reduce corrosion, assessed by measuring the proportion of free metallic lead. Achieving the correct humidity level is equally critical to ensure a consistent acid ratio in the final battery. Humidity levels are evaluated by comparing the measured value to the relative humidity (RH). Failure to meet curing or humidity targets leads to reduced product quality, including poor plate formation, electrolyte imbalance, lower charge acceptance, shorter lifespan, and diminished battery capacity.

Each heat-treatment program specifies a temperature and humidity profile, gradually ramped over time. These profiles represent the target conditions – optimal to complete curing and achieve desired humidity level – for lead plates and are established during product development. Depending on their characteristics and requirements, multiple lead plate types may be processed using the same heat-treatment program. Maintaining these profiles requires energy, and deviations – either below or above target values – can negatively impact product quality. *Figure 1* visualizes the abstracted heat-treatment process, illustrating the corresponding inputs and outputs, providing a clear representation of the overall procedure.

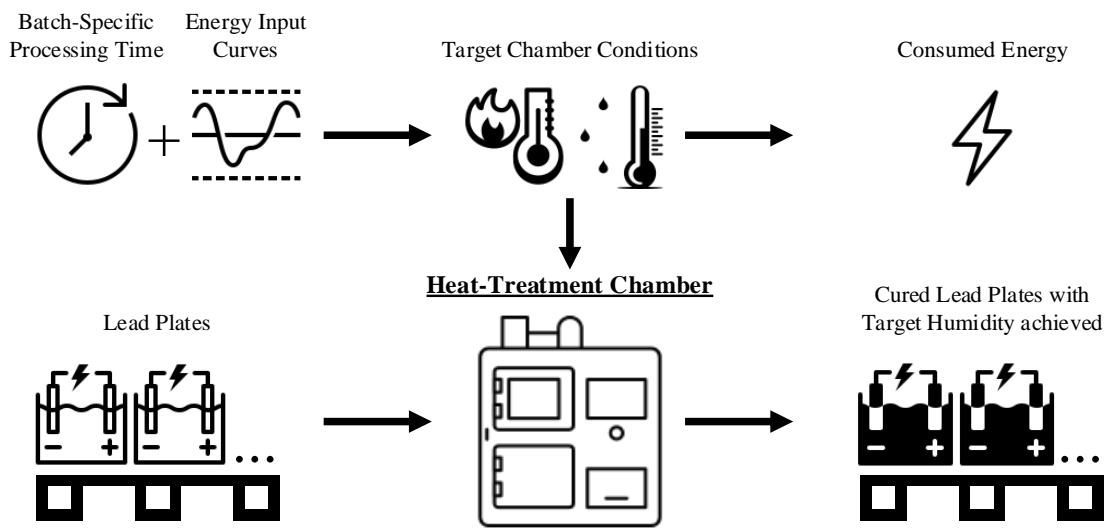

Figure 1: Abstracted Heat-Treatment Process.

*3.2 Necessitated Process Steps*

*Figure 2* provides a detailed view of the specific steps involved, with each step numbered for ease of reference. Blue-shaded steps highlight main process steps where batch-specific processing times must be determined through a stopping policy. Orange-shaded steps indicate rework processes, which incur energy penalties when chemical curing or humidity targets are not met.

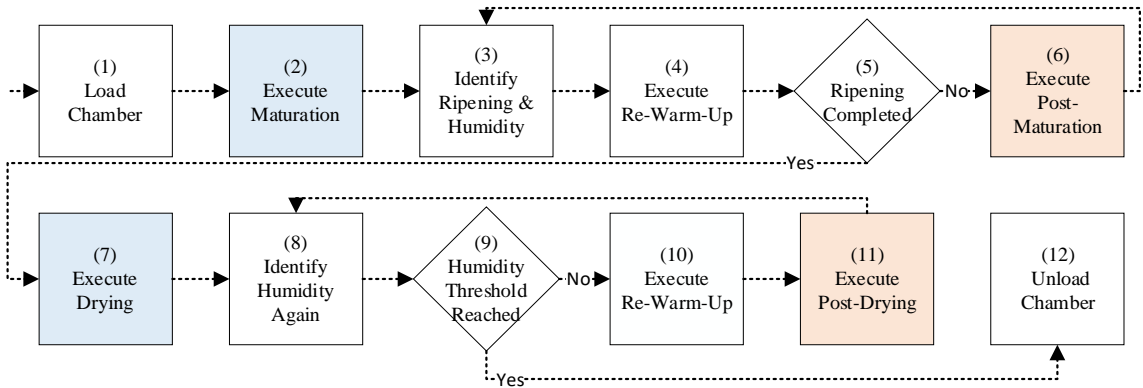

Figure 2: Heat-Treatment Process Steps.

The process starts with a forklift operator loading batched and buffered pallets of lead plates into the chamber, consuming energy during the loading phase (1). After loading, the maturation begins (2), with the chamber ramping up temperature and humidity to target levels – this fist ramp-up acts as a warm-up phase. Except for loading and unloading, all steps have no variability in the time execution; the realized processing times match the planned ones. Once the planned processing time elapses, the chamber is opened (3) for inspection. In *Current Baseline Practice*, this decision is simply taken after the planned processing time elapses. Following inspection, the chamber is re-warmed to compensate for any temperature losses (4). Depending on the inspection results (3), rework – specifically post-maturation – may be necessary until chemical curing is completed (5). During post-maturation (6), predefined planned processing times are reapplied; this may occur in multiple cycles for a single batch. As such, overly long maturation times in (2) cause unnecessary energy consumption, while too short durations may result in repeated post-maturation cycles (6) along with chamber re-warm-ups (4). After successful curing, drying begins (7), following a similar sequence. Once the drying time elapses, humidity is measured (8) and compared to a threshold humidity level (9). Depending on the results, re-warm-up (10) and post-drying (11) are repeated as needed. Again, balancing overly long drying times (7) against the need for additional post-drying cycles (11) and chamber re-warm-ups (10) is crucial. The process ends with unloading (12), which requires no energy, as the chamber is deactivated.

*3.3 Energy Input Curves*

To analyse energy consumption over time, *Figure 3* visualizes the target temperature (°C) and humidity (g/m³), along with the corresponding energy inputs (kWh), for each of the four heat-treatment programs at the case company. Different programs are necessary because various lead plate types require specific chamber conditions over time to complete maturation and achieve targeted humidity.

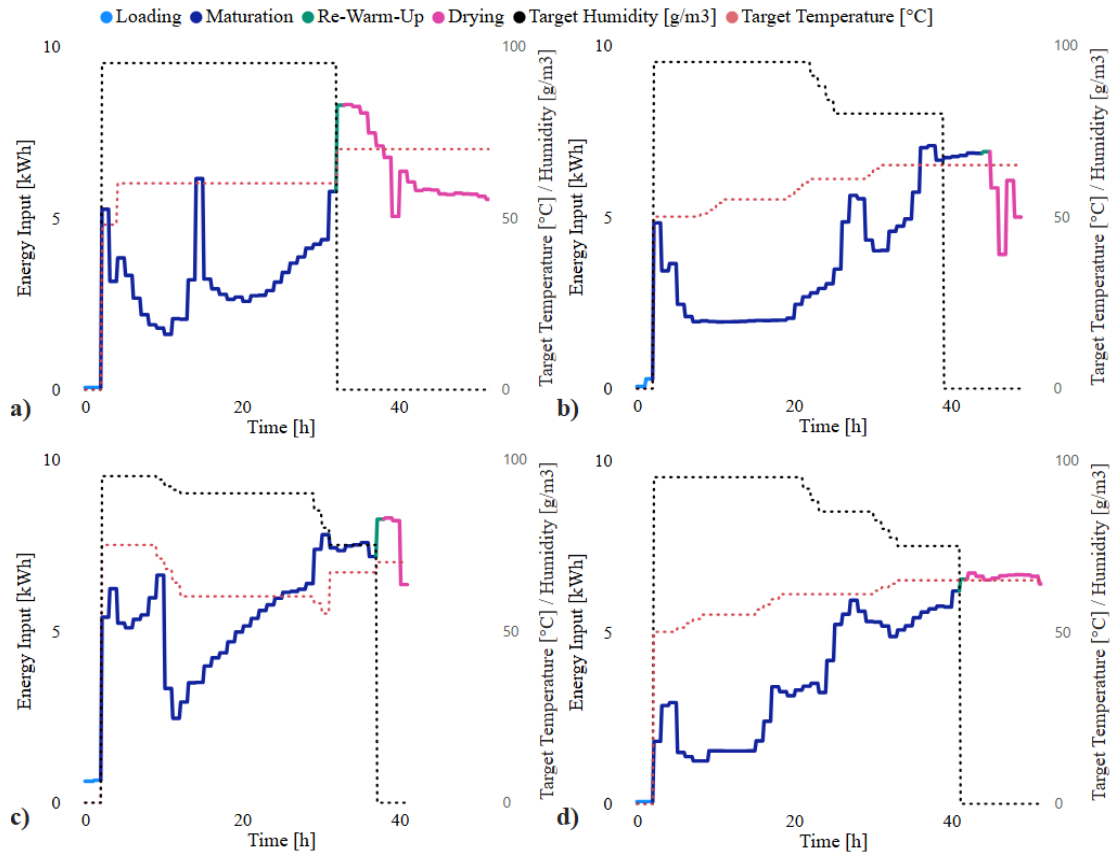

Figure 3. Energy Input Curves Heat-Treatment Program:
a) Negative; b) Positive; c) Positive-with-Vaporization; d) Start/Stop.

The data is based on measurements from batches that completed without rework. Axes are equally scaled for comparison; therefore some values are missing for programs with shorter planned processing times. All programs show a sharp initial energy peak to reach temperature and humidity, effectively serving as warm-up. During early maturation, heat from chemical reactions reduces energy input, but energy demand rises again as target temperature increases and self-heating decreases. Following the inspection, the necessary re-warm-up after opening the chamber adds another energy spike. During drying, energy increases further as temperature rises and humidity is reduced to zero.

Among the programs, the Positive-with-Vaporization (*Figure 3c*) shows the highest energy consumption due to its high target temperature and steam generation requirements. In contrast, the Start/Stop program (*Figure 3d*) starts with lower energy, reflecting its reduced initial temperature. The Negative program (*Figure 3a*) follows the simplest profile, with a linear ramp-up in temperature and humidity, whereas the other programs use more complex, multi-ramp curves.

## *3.4 Problem Formulation*

The heat-treatment problem is structured as two independent stopping problems – one for completing chemical curing, and the other for achieving the target humidity level. Each consists of a main process step (maturation or drying) and a rework step (post-maturation or post-drying), as shown in *Figure 2*. To satisfy process quality, the total energy input – provided at the respective main process step and any corresponding rework – must meet or exceed a batch-specific unknown minimum required energy. This value depends on stochastic factors such as moisture content, paste thickness, and thermal losses, and is not directly measurable. The energy input can be considered as a deterministic function being the amount of heat transferred into the process controllable by heat-treatment chamber. Specifically, given the function shown in *Figure 3* and the applied processing time, the energy input is determined by summing up these curves over the applied processing time. The current analysis focuses on a single chamber, assuming that energy during loading is unaffected by planning. Future work will address multi-chamber scheduling. By applying the stopping logic separately for curing and drying, the optimization problem is simplified, resulting in a formulation that is broadly applicable to stochastic production systems aiming to minimize energy input across both main and rework phases.

The stopping problem seeks the batch-specific processing time *T* that minimizes the total energy input – the sum of energy consumed during the main process step *m*, re-warm-up *r*, and rework *g* – as in:

$$\min_{\{T\}} \left( \varphi^c(T) + \sum_{j=1}^{J} \left( \psi^r + \psi^g \right) \right) \tag{1}$$

Where, $\varphi^c(T)$ represents the cumulated energy input for the main process step *m* at the applied processing time *T* (equal to the realized time). Meanwhile, $\psi^r$ and $\psi^g$ denote constant energy inputs for re-warm-up *r* and rework *g*, respectively. Because $\psi^r$ is fixed by equipment settings and $\psi^g$ is dictated by process requirements (i.e., fixed planned processing time), neither constant can be altered through planning decisions. Rework cycles *J* occur only if the cumulated energy input $\varphi^m(T)$ provided at the main step, fails to reach the minimum required energy $\Phi$, as shown in *Figure 2* and expressed by:

$$\varphi^c(T) + \sum_{j=1}^{J} \psi^g \geq \Phi \tag{2}$$

If *Constraint (2)* is already satisfied at the end of the critical step, $J=0$ and the rework term in *Objective Function (1)* is zero. Otherwise, each additional cycle $j$ adds a fixed re-warm-up $\psi^r$ plus rework energy $\psi^g$ until $\Phi$ is met. Notice that when $T$ is determined then the value of $J$ is known, or vice versa.

| Notation | Description |
| --- | --- |
| $n$ | Iteration Counter |
| $\tau$ | Constant Time Interval between Iterations $n$ |
| $w$ | Warm-up |
| $m$ | Main Process Step $m$ |
| $r$ | Re-Warm-Up |
| $g$ | Rework |
| $T$ | Batch-Specific Processing Time at Main Process Step $m$ |
| $N$ | Stopping Iteration at Main Process Step $m$ |
| $J$ | Number of Rework Cycles |
| $\psi^x$ | Constant Energy Input with $x=r$ for Re-Warm-Up $x=g$ for Rework |
| $\psi_n^m$ | Energy Input at Main Process Step $m$ at Iteration $n$ |
| $\varphi^c(\cdot)$ | Cumulated Energy Input at Main Process Step $m$ for Processing Time $T$ or Stopping Iterations $N$ multiplied by Constant Time Interval $\tau$ |
| $\varphi_n^e$ | Bayesian Energy Estimate at Iteration $n$ |
| $\varphi_n^t$ | Energy-Safety Threshold at Iteration $n$ |
| $\Phi$ | Batch-Specific Minimum Required Energy (True and Hidden) |
| $K_n^{\varphi^t}$ | Iteration Counter to Reach Safety Threshold $\varphi^t$ at Iteration $n$ |
| $L^{E[\Phi]}$ | Iteration Counter to Reach Expected Minimum Required Energy $E[\Phi]$ |
| $S_n$ | State at Iteration $n$ |
| $\upsilon_n$ | Sensor Reading at Iteration $n$ |
| $\eta_n$ | Sensor Refinement at Iteration $n$ |
| $p_n$ | Expected Remaining Batch-Specific Processing Time at Iteration $n$ |
| $a_n$ | Control action {continue / terminate} at iteration $n$ |
| $\Delta_n$ | Sensor Deviation – Minimum Required Energy $\Phi$ and Sensor Reading $\upsilon_n$ at Iteration $n$ |
| $\beta$ | Energy-Safety Parameter |
| $\alpha_n^{\upsilon}$ | Sensor Distortion (Coefficient of Variation) for Iteration $n$ |
| $\alpha^{E[\Phi]}$ | Coefficient of Variation of Expected Minimum Required Energy $E[\Phi]$ |
| $\sigma^{\varphi^e}$ | Standard Deviation of Bayesian Energy Estimate $\varphi^e$ |
| $E[\ ]$ | Expected Value |
| $F^{-1}$ | Inverse of Cumulative Distribution Function (CDF) |

Table 1: Notations.

## 4. Simulation-Based Approach

The minimum required energy $\Phi$ is unknown and stochastic, so processing time $T$ cannot be chosen directly from *Constraint (2)*. We therefore develop a simulation-based stopping policy that adjusts expected remaining processing time and decides when to terminate. *Figure 4* shows the abstracted procedure performed iteratively during the main step *m*.

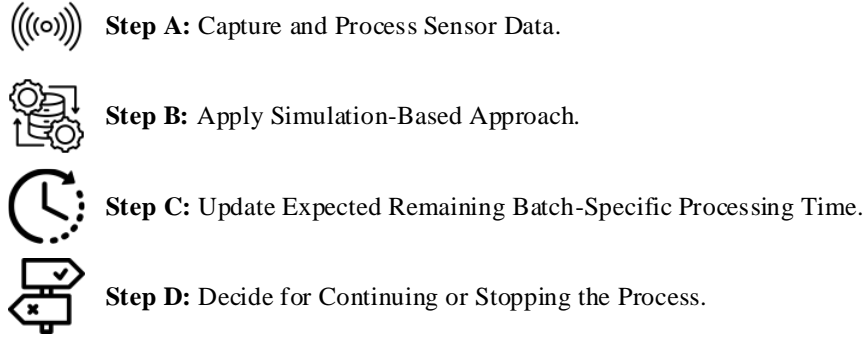

**Step A:** Capture and Process Sensor Data.

**Step B:** Apply Simulation-Based Approach.

**Step C:** Update Expected Remaining Batch-Specific Processing Time.

**Step D:** Decide for Continuing or Stopping the Process.

Figure 4. Abstracted Procedure Simulation-Based Approach.

*4.1 Discretising the Stopping Problem*

We divide processing time $T$ into iterations of fixed length $\tau$. Let $n$ denote the current iteration and $N$ the iteration at which the main step halts. Thus, cumulated energy input $\varphi^c$ at the main step *m* is:

$$\varphi^c(T) = \varphi^c(N\tau) = \sum_{n=0}^{N} \psi_n^m \tag{3}$$

where $\psi_n^m$ is the energy input during iteration $n$. With time discretised, the continuous energy input curves in *Figure 3* can be integrated by applying integral approximation to calculate $\varphi^c(N\tau)$. The stopping problem – selecting the optimal stopping time – reduces to finding $N$ that minimises the total energy in *Objective Function (1)*. Our *SBA* defines $N$ specifically for each batch, while any required cycles $J$ are added afterwards repeatedly until *Constraint (2)* is satisfied.

To link energy accumulation to control decisions we invert the cumulative energy input curves (obtained by cumulating values from *Figure 3*). This inversion yields two separate counters. First, the iterations needed to provide the expected minimum required energy $E[\Phi]$:

$$L^{E[\Phi]} = \min_{k} \quad \text{with} \quad \varphi^c(k\tau) \geq E[\Phi] \tag{4}$$

and second, the iterations required – at current $n$ – to reach the energy-safety threshold $\varphi_n^t$:

$$K_n^{\varphi^t} = \min_k \quad \text{with} \quad \varphi_n^c(k\tau) \geq \varphi_n^t \qquad (5)$$

The first counter $L^{E[\Phi]}$ later scales the sensor distortion as processing progresses, whereas the second counter $K_n^{\varphi^t}$ feeds directly into the expected remaining batch-specific processing time estimate $p_n$ and the stop/continue decision.

## *4.2 State Representation*

To determine the batch-specific stopping iteration $N$, the problem is embedded within a Markov Decision Process that is iteratively updated. At iteration $n$, the Markovian model stores the following nine-component state vector:

$$S_n = \{n \,;\, \varphi^c \,;\, \Phi \,;\, \upsilon_n \,;\, \Delta_n \,;\, \varphi_n^e \,;\, \varphi_n^t \,;\, p_n \,;\, a_n\} \qquad (6)$$

here $n$ is the iteration counter, $\varphi^c$ the cumulated energy provided up to that point *(Equation 3)*; $\Phi$ the true but hidden batch-specific minimum energy required to satisfy the target feature (i.e., complete ripening or achieve target humidity); $\upsilon_n$ the current sensor reading; $\Delta_n = \Phi - \upsilon_n$ the resulting deviation; $\varphi_n^e$ the Bayesian energy estimate of $\Phi$; $\varphi_n^t$ the computed energy-safety threshold; $p_n$ the corresponding expected remaining batch-specific processing time to reach $\varphi_n^t$; and $a_n \in \{\text{continue, terminate}\}$ the control action.

At the start of the process $n=0$, the state vector is initialized as:

$$S_0 = \begin{cases} n = 0 \,;\, \varphi^c = 0 \,;\, \Phi \sim LogN_{Mean-CV}\left(E[\Phi], \alpha^{E[\Phi]}\right); \upsilon_0 = E[\Phi]; \Delta_0 = \Phi - \upsilon_0 \,; \\ \varphi_0^e = E[\Phi] \,;\, \varphi_0^t = F_N^{-1}\left(\beta; \varphi_0^e, \sigma^{\varphi^e}\right); p_0 = \left(K_0^{\varphi^t} - 0\right)\tau \,;\, a_0 = \text{continue} \end{cases} \qquad (7)$$

here the true but hidden batch-specific minimum energy $\Phi$ is sampled from a log-normal distribution with the expected value $E[\Phi]$ and the corresponding coefficient of variation (CV) $\alpha^{E[\Phi]}$ obtained from historical data (i.e., long-term fluctuations of $\Phi$). Both the sensor reading $\upsilon_0$ and the Bayesian energy estimate $\varphi_0^e$ are seeded with the expected value $E[\Phi]$. The energy-safety threshold $\varphi_0^t$ is computed using a Gaussian inverse cumulative distribution function (CDF) evaluated at $\beta$. The corresponding remaining processing time is derived from the number of iterations required to reach the threshold $L^{E[\Phi]}$ via the cumulative energy curve *(Equation 4)*.

*Figure 5* illustrates this initial state: the true required energy $\Phi = 110\text{kWh}$ (red line) is sampled from a log-normal distribution with $Mean = 100\text{kWh}, CV = 0.2$ (blue curve, note logarithms of these inputs). Thus, both sensor reading $\upsilon_0$ and Bayesian estimate $\varphi_0^e$ equal 100kWh (purple and orange dotted lines, overlapping). Consequently, the resulting deviation $\Delta_0$ is 10kWh (black line). The threshold $\varphi_0^t = 132.9\text{kWh}$, derived at $\beta = 0.95$ (dark blue dotted line).

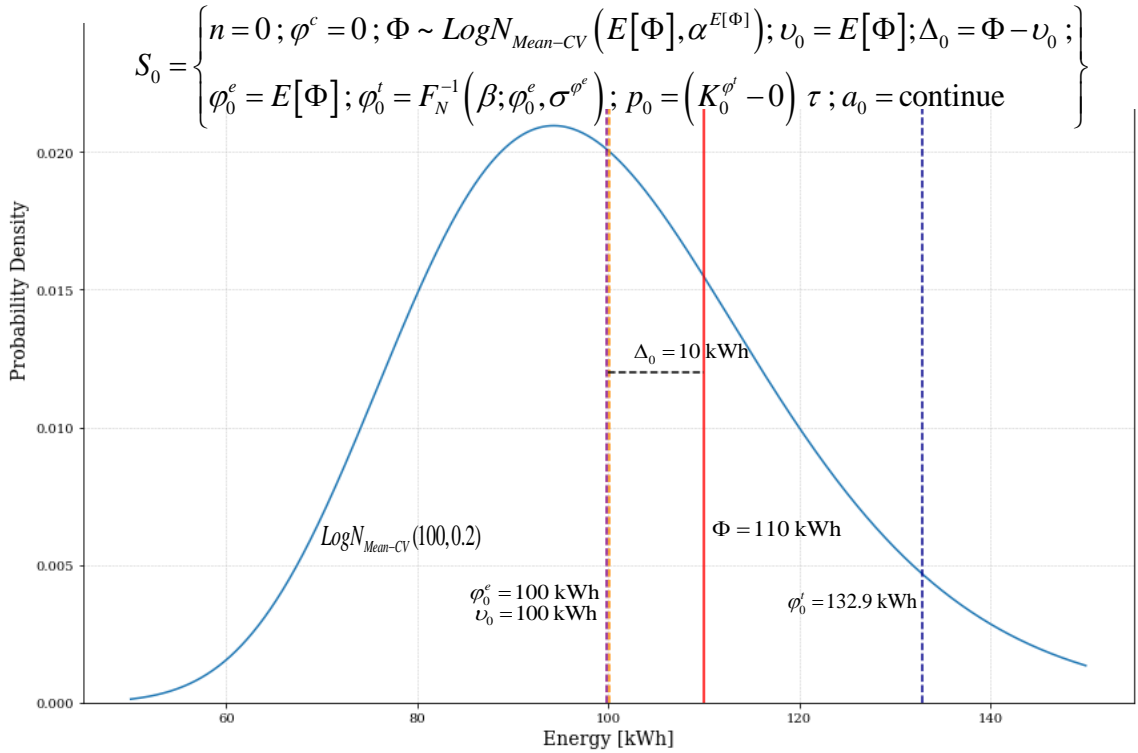

Figure 5: Visualizing Initial Iteration State.

### *4.3 Iterative Sensor-Driven Markovian Loop with Bayesian Update*

At each iteration *n*, the Markov model generates a simulated sensor reading $\upsilon_n$. The next reading $\upsilon_{n+1}$ is obtained by correcting the current $\Delta_n = \Phi - \upsilon_n$ with a stochastic refinement:

$$\upsilon_{n+1} = \upsilon_n + \eta_{n+1}, \qquad \eta_{n+1} \sim N(\Delta_n, (\alpha_n^\upsilon \upsilon_n)^2) \qquad (8)$$

Where $\alpha_n^\upsilon$ is the iteration-dependent sensor distortion, i.e., coefficient of variation. Because the normal mean equals $\Delta_n$, the correction term $\upsilon_{n+1}$ automatically steers the reading toward the hidden target $\Phi$. Thus, if $\upsilon_n$ is too low (i.e., $\Delta_n > 0$), positive draws are more likely, pushing the next reading upward. Conversely, if $\upsilon_n$ is too high (i.e., $\Delta_n < 0$), negative draws dominate, pulling the next reading downward. The variance $(\alpha_n^\upsilon \upsilon_n)^2$ reflects the empirical observation that measurement noise scales with signal magnitude.

Uncertainty should shrink as processing progresses and the batch approaches its expected energy requirement $E[\Phi]$. We therefore let the sensor distortion (used as CV) decline linearly from an initial value $\alpha_0^\upsilon$ to a floor value $\alpha_{L^{E[\Phi]}}^\upsilon$:

$$\alpha_n^\upsilon = \max\left(\alpha_{L^{E[\Phi]}}^\upsilon, \alpha_0^\upsilon - \left(\frac{\alpha_0^\upsilon - \alpha_{L^{E[\Phi]}}^\upsilon}{L^{E[\Phi]}}\right) n\right) \tag{9}$$

where $L^{[\Phi]}$ *(Equation 4)* is the minimum number of iterations to reach the expected minimum required energy. Anchoring the slope to this counter ensures sensor distortion scale as processing progresses. The max operator prevents $\alpha_n^\upsilon$ from falling below the predefined floor $\alpha_{L^{E[em]}}^\upsilon$, for any $n > L^{E[\Phi]}$.

The updated sensor reading $\upsilon_{n+1}$ is combined with the previous Bayesian energy estimate $\varphi_n^e$ to form a new posterior as:

$$\varphi_{n+1}^e = \frac{\varphi_n^e + \upsilon_{n+1}}{2} \tag{10}$$

Here, $\varphi_n^e$ serves as prior and the updated sensor reading $\upsilon_{n+1}$ as likelihood. From this estimate we derive an energy-safety threshold as:

$$\varphi_{n+1}^t = F_N^{-1}\left(\beta; \varphi_{n+1}^e, \sigma^{\varphi^e}\right) \tag{11}$$

using the Gaussian inverse-CDF evaluated at the chosen risk quantile $\beta$. Here, $\sigma^{\varphi^e}$ represented the long-term SD of the estimates obtained from historical sensor data. A larger $\beta$ raises the threshold, lowering the risk of rework cycles $J$ *(Constraint 2)* at the expense of longer processing and potentially unnecessary energy use during the main step. Based on the required number of iterations to reach the energy-safety threshold $K_{n+1}^{\varphi^t}$ *(Equation 5)*, the remaining batch-specific processing time $p_{n+1}$ is:

$$p_{n+1} = \left(K_{n+1}^{\varphi^t} - (n+1)\right)\tau \tag{12}$$

With the term *(n+1)* representing the iterations already completed. The result drives the action $a_n$:

- **Continue:** If $p_{n+1} > 0$, proceed to the next iteration $n+2$.
- **Terminate:** If $p_{n+1} \leq 0$, set the stopping index $N = n+1$.

When terminated, the cumulated energy input $\varphi^c(N\tau)$ is evaluated via *Equation (3)*, an inspection verifies *Constraint (2)* and, if necessary, rework cycles are iteratively performed until the batch-specific minimum energy requirement $\Phi$ is satisfied, yielding the final rework count *J*. *Figure 6* summarises the sensor-driven stopping loop within the Markov Decision Process.

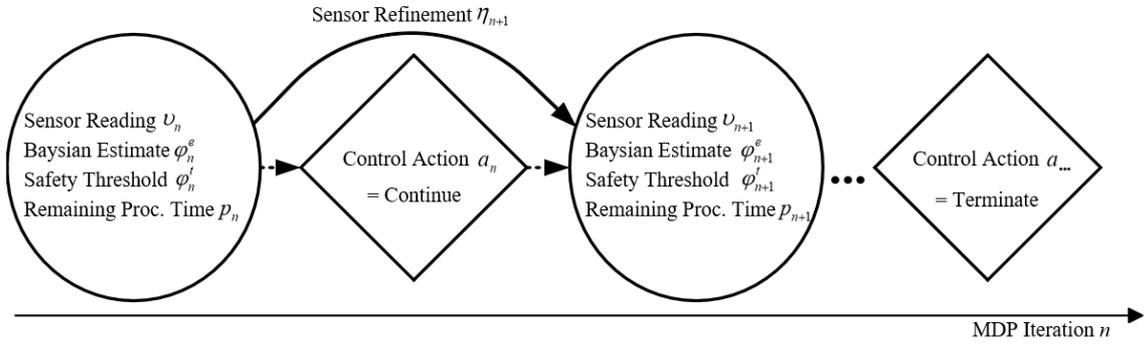

Figure 6: Abstracted Sensor-Driven Stopping Loop within Markov Decision Process.

To clarify the mechanics, *Figure 7* walks through the computation of the expected remaining batch-specific processing time $p_n$ for a concrete case. At iteration *n=40* (and $\tau = 1$ time unit), the controller obtains the cumulative energy input $\varphi^c(40)$ (accumulated maturation curve in *Figure 3a*) and compounds the energy-safety threshold $\varphi^t_{40}$ on the same curve *(Equation 11)*. Inverting the curve yields the iteration counter $K^{\varphi^t}_{40}$ *(Equation 5)*, which specifies how many further iterations are required to reach the threshold. The difference between $K^{\varphi^t}_{40} - 40$ therefore gives the additional iterations still needed; multiplying this count by $\tau$ yields the remaining processing time $p_{40} = 40\,[\text{TU}]$.

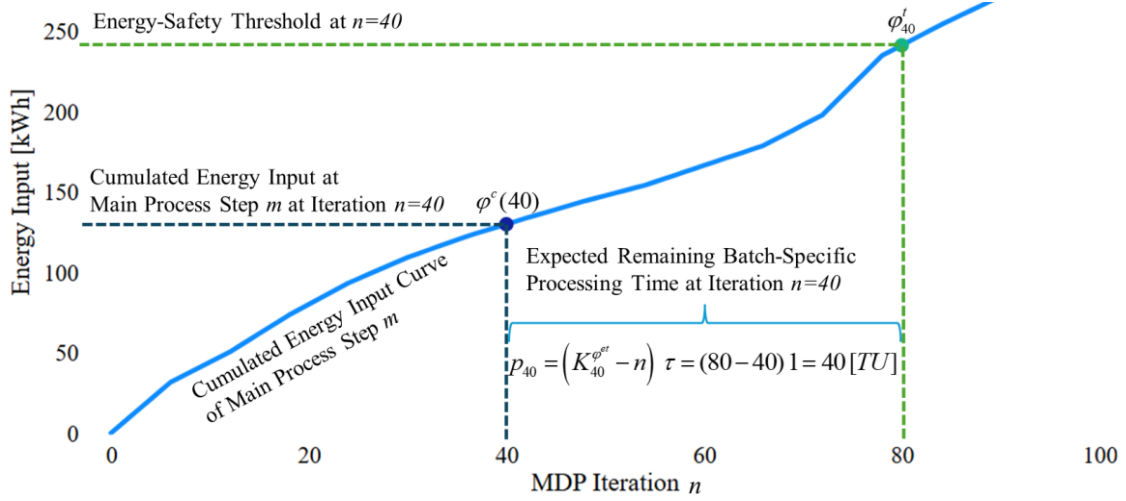

Figure 7: Calculation of the Expected Remaining Batch-Specific Processing Time.

## *4.5 Example Run of Simulation-Based Approach*

To illustrate how the loop just described unfolds in the simulation, *Figure 8* tracks the state-vector components defined in *Equation (6)* over successive iterations for one representative batch processed with the negative programme. For this program, the expected minimum required energy is $E[\Phi] = 308.84$ kWh with a CV of $\alpha^{E[\Phi]} = 0.30$. The batch we follow happens to draw a lower-than-average requirement $\Phi = 253.24$ kWh. We set an energy-safety parameter $\beta = 0.95$. Initial sensor distortion equals $\alpha_0^v = 0.2$ and the floor value $\alpha_{L^{E[\Phi]}}^v = 0.02$.

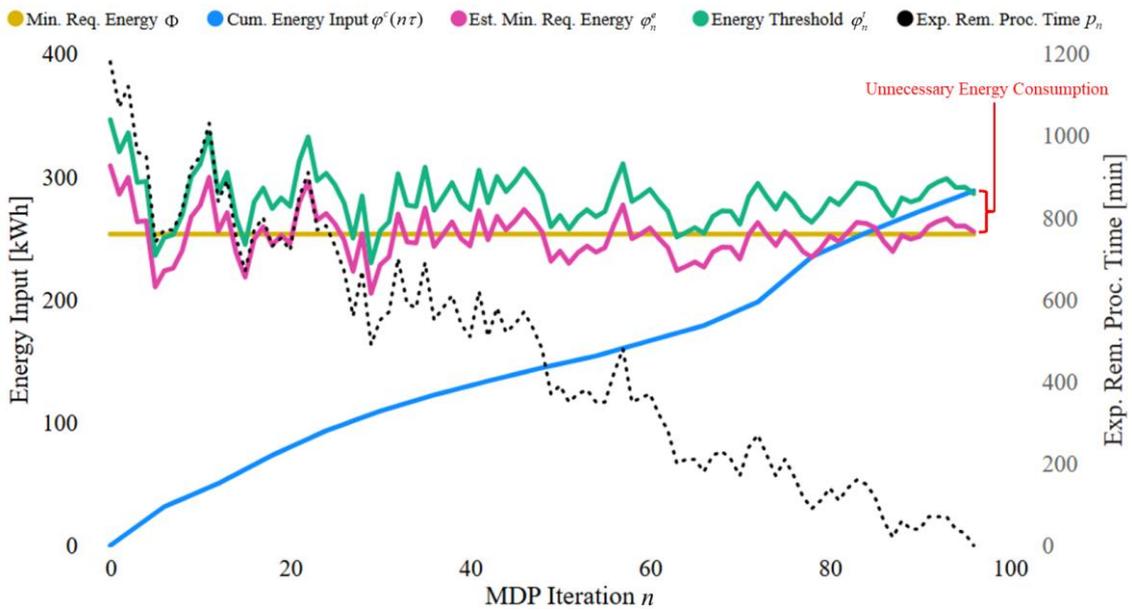

Figure 8: Changes of State Attributes over Iteration.

At early phase *(n<30)* with distortions $\alpha_n^v$ still high, sensor readings $v_n$ are noisy and both Bayesian energy estimate $\varphi_n^e$ and energy-safety threshold $\varphi_n^t$ fluctuate widely. As *n* increases $\alpha_n^v$ decrease linearly toward the floor values *(Equation 9)*; noise shrinks and $\varphi_n^e$ converges towards the hidden target $\Phi$. Because $\beta = 0.95$ is conservative, $\varphi_n^t$ remains well above the true requirement $\Phi$, increasing the batch-specific processing time $p_n$ unnecessarily long, leading to excessive energy consumption.

## 5. Simulation Model

To evaluate the energy reduction potential of the proposed *SBA*, we developed an agent-based discrete-event simulation model that mirrors the heat-treatment process of the case company. Building on the preliminary study by Bokor and Altendorfer (2023a), which relied on a simplified process and sensor approximation with incomplete company data – this version mirrors realistic sensor behavior and a case-oriented implementation. Most technological details (programs, step sequence, energy curves, stopping logic) are covered in Sections 3 and 4, so this section focuses on modelling choices, stochastic assumptions, and configuration. The simulation is a multi-item, multi-stage agent model centered on a single electrically powered 32-pallet chamber (one of multiple operational chambers at the case company) and the pre-production lines. Real-world uncertainty is introduced via stochastic order arrivals, lot sizes, loading/unloading times and batch-specific minimum energy requirements. *Table 2* – derived from 2024 operational data – summarizes the parametrization's, grouped into "Production Order" for 21 lead plate types and "Heat-Treatment Process" for the four heat treatment programs. Process step numbering follows *Figure 2*. Parameters varied later in the numerical study are reported only in *Section 6* to avoid redundancy.

| Production Order | Mean | CV |
|---|---|---|
| Inter-Arrival Time [Orders / Working Day] | 0.1 | 0.2 |
| Lead Plate Types | 0.14; 0.05; 0.01; 0.01; 0.03; 0.01; 0.01; 0.01; 0.14; 0.11; 0.06; 0.08; 0.03; 0.01; 0.06; 0.01; 0.03; 0.13; 0.01; 0.05; 0.01 | - |
| Lot Size [Palette] | 153; 76; 49; 29; 48; 39; 28; 37; 159; 119; 77; 88; 49; 35; 76; 32; 62; 128; 32; 59; 44 | 0.33; 0.47; 0.43; 0.31; 0.52; 0.43; 0.43; 0.00; 0.31; 0.36; 0.47; 0.42; 0.45; 0.37; 0.36; 0.28; 0.32; 0.39; 0.32; 0.46; 0.00 |
| **Heat-Treatment Process** | **Mean** | **CV** |
| (1) Loading [Min. / Palette] | 4 | 0.2 |
| (12) Unloading [Min. / Palette] | 4 | 0.2 |
| (2) Maturation based on Time $T$ / Iteration $N$ | Tested in Numerical Study | - |
| (4) Re-Warm-Up Post-Maturation [Min. / Batch] | 60 | - |
| (6) Post-Maturation [Min. / Batch] | Tested in Numerical Study | - |
| (5) Exp. Min. Energy Curing $E[\Phi]$ [kWh] | 308.84; 641.36; 934.08; 249.14 | Estimated; Tested in Numerical Study |
| (7) Drying based on Time $T$ / Iteration $N$ | Tested in Numerical Study | - |
| (10) Re-Warm-Up Post-Drying [Min. / Batch] | 60 | - |
| (11) Post-Drying [Min. / Batch] | Tested in Numerical Study | - |
| (9) Exp. Min. Energy Humidity $E[\Phi]$ [kWh] | 604.51; 582.58; 368.72; 1,084.27 | Estimated; Tested in Numerical Study |

Table 2: Simulation Model Parametrization derived from Company Data.

## *5.1 Production Order Modelling*

Orders arrive according to a log-normal inter-arrival time (mean=0.1 orders / working day, CV=0.20). The low rate ensures that the single chamber never becomes a bottleneck and ensures throughput comparable across all applied processing times. Since achieving targeted humidity span several days, and production order lot sizes often exceed chamber capacity – necessitating multiple batches per order. Each order contains one of 21 plate types (selection probabilities in Table 2) and a lot size drawn from a plate-specific log-normal distribution (mean / CV in Table 2). Lots exceeding the 32-pallet capacity are split into sequential batches. Pre-production (i.e., casting and pasting of lead plates with lead oxide, along with setup operations when changing plate types) is simplified to a stochastic delay, as no scheduling decision is applied. Loading and unloading proceed batch-wise and strictly FIFO, so chamber is never shared by different order numbers.

## *5.2 Heat-Treatment Process Modelling*

Every batch runs one of the four programmes shown in *Figure 3*. Energy input follows the programme's curve. If the process time extends beyond measured data, the curve is extrapolated linearly according to the step's end. Post-maturation and post-drying rework consume the energy value recorded at the end of the respective main step.

There is no variability in the execution of any process step (i.e., the realized processing times matches target), except for loading (1) and unloading (12), which depend on forklift performance; each modelled as a log-normal random time (mean=4 min / palette, CV=0.2 per pallet). Re-warm-up after post-maturation (4) and post-drying (10) are fixed by equipment settings at 60 min per batch. Process times for maturation (3) and drying (7) are either set to the planned processing times or generated dynamically by the *SBA*. By contrast, post-maturation (6) and post-drying (11) always follow their planned processing times, which are varied in the numerical study.

Batch-specific minimum energies for curing (5) and humidity (9) are program-specific log-normal variables that capture variability in moisture, paste thickness, and chamber losses, thereby dictating rework likelihood. As direct parameter measurement is impractical, means (shown in *Table 2*) are derived from metallic lead inspection data and different levels of uncertainty (i.e., CV) are stress-tested in the numerical study to assess their impact on energy consumption and parametrization. Detailed data and calculations are provided in the *Appendix*. The long-term SD of the estimates $\sigma^{\varphi^e}$, required to compound the energy-safety threshold $\varphi_n^t$ (*Equation 11*), is calculated based on data obtained during the simulation warm-up phase, separately for curing and humidity.

The simulation model declares curing (3) or drying (9) complete when the stochastic minimum energy draw is met or exceeded by the deterministic cumulative input provided during maturation (plus any post-maturation) or drying (plus any post-drying).

## 6. Numerical Study

We conduct a comprehensive full-factorial study to quantify the energy reduction potential achieved by the *SBA* with batch-specific processing times. For benchmarking, we include an *OPT* (*Optimized Planned Processing Times*) scenario, which applies a single energy factor to the expected requirement and uses the resulting planned processing times uniformly across all batches. *Table 3* lists the investigated parameters for both *SBA* and *OPT* together with their ranges selected from preliminary runs.

| Environmental Parameters | Min | Max | Step Size | # Values |
|---|---|---|---|---|
| CV Min. Required Energy Factor $\alpha^{E[em]}$ | 0.15 | 0.45 | 0.15 | 3 |
| Sensor Distortion $\alpha_0^u$ at Iteration $n=0$ | 0.05 | 0.50 | 0.05 | 10 |
| **Planning Parameters** | | | | |
| Processing-Energy Factor – Main Process Steps | 0.70 | 1.25 | 0.05 | 12 |
| Processing-Energy Factor – Rework Process Steps | 0.05 | 0.60 | 0.05 | 12 |
| Energy-Safety Parameter $\beta$ – Critical Process Steps | 0.40 | 0.95 | 0.05 | 12 |
| # Iterations – *OPT* (*Optimized Planned Processing Times Approach*) | | | | 432 |
| # Iterations – *SBA* (*Simulation-Based Approach*) | | | | 4,320 |
| **Total Iterations** | | | | **4,752** |
| Simulation Runs (20 Replications per Iteration) | | | | 95,040 |

Table 3: Investigated Parameters and Values at the Numerical Study.

To investigate environmental scenarios, we vary process uncertainty and sensor distortion. Uncertainty is examined at three CV levels – low, medium, high. The medium level (CV=0.30) reflects the weighted average of 0.31 across the four heat-treatment programmes (see *Appendix*) and represents the case company's current operational condition. Sensor distortion is explored at ten different initial values $\alpha_0^\upsilon$, while the floor value is fixed to $\alpha_{L^{E[\Phi]}}^\upsilon = 0.02$, once the expected energy has been provided. Identical scenarios apply to both curing and drying.

For *OPT*, we test twelve processing-energy factors for the main process steps (maturation and drying) and their rework steps (post-maturation and post-drying). Each factor specifies the cumulated energy input as a multiple of the expected minimum. For instance, a factor of 1.10 on the Negative programme (mean=390 kWh) corresponds to 429 kWh and, according to the cumulative curve in *Figure 3*, a planned processing time of about 1,000 min. For *SBA* those main step factors are replaced by twelve energy-safety parameters β. Each β sets the risk quantile and therefore the threshold $\varphi_n^t$ (*Equation 11*) and batch-specific processing time $p_n$ (*Equation 12*). Rework steps for *SBA* keep the same twelve energy factors used as for *OPT*. Parameters (i.e. factors and β) are applied identically to both main steps and to both rework steps.

The full factorial design generates 432 *OPT* iterations and 4,320 *SBA* iterations; each is replicated 20 times to account for stochastic influences, resulting in 95,040 runs. Each run spans four production years (first year warm-up) and finishes in approximately 30 seconds (real time).

Parameter combinations are generated in Spyder/Python, distributed across 21 networked PCs for parallel computing in AnyLogic 8.8.6, and the results are written back to a PostgreSQL database for analysis.

## 7. Numerical Results

To comprehensively evaluate the energy reduction potential of the *SBA* under varying environmental scenarios (i.e., process uncertainty and sensor distortion) we systematically analyse the entire heat-treatment process. Energy input for loading is excluded throughout, as it remains unaffected by the decisions currently under investigation. Results of the *OPT* appear first, followed by the *SBA* and the impact of sensor accuracy. Next, *SBA* is benchmarked against *OPT,* the *Current Baseline Practice* and an *Ideal Scenario*, where energy requirements are precisely known, and processing times are optimally set. Finally, energy consumption is converted into monetary terms and combined with inspection-labour costs to reveal the trade-off between energy and personnel expenses.

*7.1 Analysis Optimized Planned Processing Times Approach*

At *OPT*, fixed planned processing times are applied uniformly across all batches at the main steps maturation and drying. *Figure 9* illustrates the impact of different processing-energy factors at these steps across the three levels of process uncertainty, represented by the CV of the minimum required energy. The analysis considers the optimal processing-energy factor for rework in each scenario. The primary y-axis shows the minimum average cumulative energy input per batch (in kWh), while the secondary y-axis displays the rework ratio, defined as the proportion of batches needing at least one rework step relative to the total throughput. The *Current Baseline Practice* (factor=1.2 at CV=0.3) is highlighted in red (assuming optimal rework processing-energy factor).

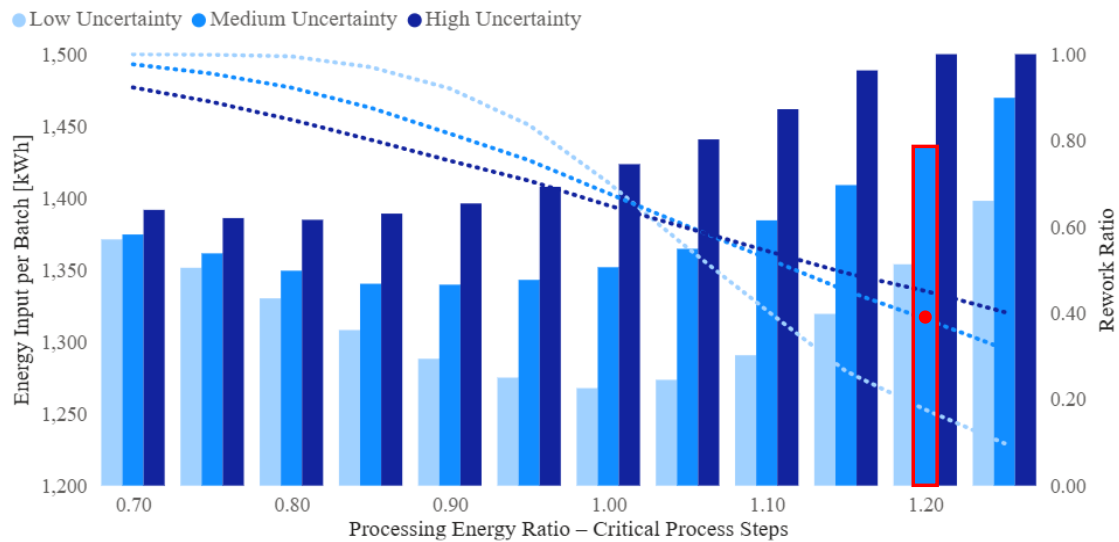

Figure 9: Impact of Planned Processing Times at Main Process Steps.

Higher process uncertainty clearly leads to higher energy consumption. At low uncertainty, a processing-energy factor of 1.00 achieves the lowest energy input (i.e., providing expected energy requirement). Under medium and high uncertainty, however, energy input is minimized by shortening main process steps and reallocating energy toward rework. Optimal factors are approximately 0.90 for medium uncertainty and 0.80 for high uncertainty. Shorter planned processing times at critical steps are thus generally more energy-efficient due to the relatively low energy penalty for rework (only a 60-minute warm-up). The rework ratio curve flattens at lower processing-energy factors, particularly at higher uncertainty, as many batches with low energy requirements avoid rework altogether. At a 1.00 factor, around 65–70% of batches require rework because curing and drying represent two independent stopping problems, each with its own stochastically drawn minimum energy requirement. Overall, the company's current practice is overly conservative, and substantial energy savings can be achieved by optimizing planned processing times – consistent with earlier findings by Bokor and Altendorfer (2023b), who investigated a simplified setting of the same company with incomplete data.

*7.2 Analysis Simulation-Based Approach*

*Figure 10* presents a similar analysis for the *SBA*, examining the impact of different energy-safety parameters *β*, which sets the risk quantile and thus the threshold $\varphi_n^t$ (*Equation 11*) and batch-specific processing time $p_n$ (*Equation 12*). The analysis spans the three uncertainty levels, selecting the optimal processing-energy factor for rework and assuming an initial distortion level $\alpha_0^v = 0.20$ to reflect standard sensor accuracy. The primary y-axis shows the minimum average cumulative energy input per batch, while the secondary y-axis displays the corresponding rework ratio (identical to *Figure 10*).

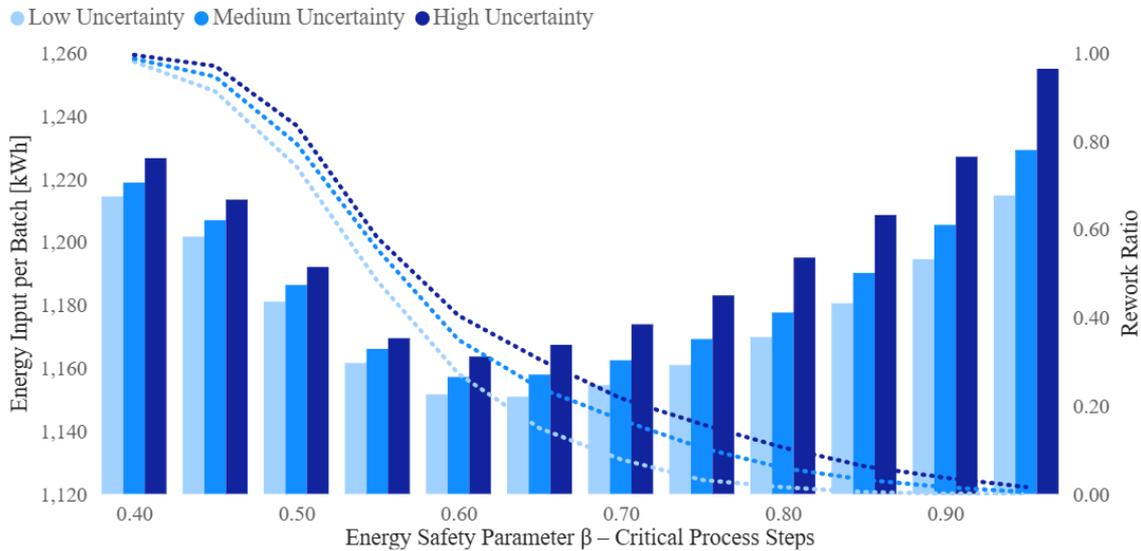

Figure 10: Impact of Energy Safety Parameters at Main Process Steps.

Results show higher uncertainty calls for slightly lower *β* values to minimize energy input. Optimal *β* values are around 0.65 at low uncertainty and around 0.60 at medium/high uncertainty. Total energy consumption remains relatively stable across tested *β* values due to the flexible redistribution of energy between main steps and rework.

Lower *β* reduces initial energy but increases rework frequency, and vice versa. Additionally, greater uncertainty consistently leads to higher rework ratios, as precise estimation becomes more difficult.

## 7.3 Impact of Sensor Accuracy

To investigate specifically how sensor accuracy affects SBA performance, *Figure 11* analyses the impact of varying initial sensor distortions $\alpha_0^v$ across the three uncertainty levels. Again, optimal processing-energy factors for rework are selected in each scenario. The primary y-axis displays the minimum average cumulative energy input per batch, while the secondary y-axis indicates the respective optimal *β*.

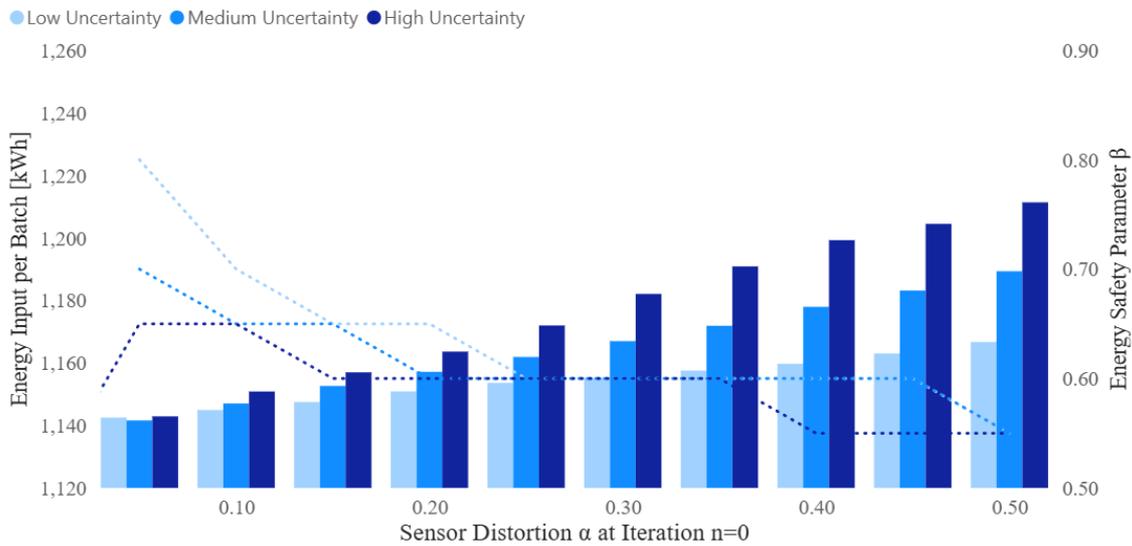

Figure 11: Impact of Sensor Distortion at Main Process Steps.

Results confirm that higher sensor distortion (lower accuracy) increases batch energy consumption due to reduced estimation reliability, but this effect is moderate. Similar energy totals can still be achieved across a range of sensor distortions by adjusting *β* accordingly. More accurate sensors (low distortion) allow higher *β* values, whereas greater distortion necessitates lower *β*, shifting more energy towards rework to handle increased uncertainty. This emphasizes the importance of balancing sensor accuracy and safety margins to ensure robust and energy-efficient process control.

## 7.4 Comparison and Benchmarking

*Table 4* summarizes minimum average cumulative energy inputs for the *Current Baseline Practice*, *OPT*, *SBA*, and *Ideal Scenario* across three levels of process uncertainty. The *Current Baseline Practice* is used as reference (100%). SBA is evaluated for three initial sensor distortion levels $\alpha_0^v = \{0.05\,;0.20\,;0.50\}$ – selecting optimal $\beta$ values each time. Again, the optimal processing-energy factor for rework is applied in each scenario. Confidence intervals are calculated at α=0.05 and α=0.01. Significant energy reductions by *SBA* compared to *OPT* are indicated by (*) for α=0.05 and (**) for α=0.01. Situations where SBA performs not significantly worse than the *Ideal Scenario* are marked with (!), based on α=0.01. No such equivalence is observed at α=0.05, indicating that SBA's performance falls outside the narrower 95 % interval, but remains within the broader 99 % range.

| **Process Uncertainty** *(CV Min. Energy)* | | **Low** | | **Medium** | | **High** | |
|---|---|---|---|---|---|---|---|
| | | Energy [kWh] | Relative Input | Energy [kWh] | Relative Input | Energy [kWh] | Relative Input |
| Current Baseline Practice | | 1,353.91 | 100% | 1,435.30 | 100% | 1,515.58 | 100% |
| OPT | | 1,267.78 | 94% | 1,339.73 | 93% | 1,384.86 | 91% |
| SBA | $\alpha_0^v = 0.05$ | ! 1,142.51 ** | 84% | ! 1,141.60 ** | 80% | ! 1,142.89 ** | 75% |
| | $\alpha_0^v = 0.20$ | 1,150.86 ** | 85% | 1,157.13 ** | 81% | 1,163.61 ** | 77% |
| | $\alpha_0^v = 0.55$ | 1,166.68 ** | 86% | 1,189.31 ** | 83% | 1,211.35 ** | 80% |
| Ideal Scenario | | 1,117.08 | 83% | 1,117.08 | 78% | 1,117.08 | 74% |

Table 4: Comparison Minimum Average Cumulative Energy Input Per Batch.

The results underline both the efficiency and robustness of the SBA. Across all environmental conditions, *SBA* significantly outperforms *OPT* even at strict α=0.01. Under conditions of low sensor distortion, *SBA* even matches the Ideal Scenario statistically at α=0.01. Relative *to Current Baseline Practice*, *SBA* requires only 84%–86% of energy at low process uncertainty, dropping to as low as 75% under high process uncertainty, representing substantial energy reduction potential.

*Figure 12* compares *OPT* and *SBA* by examining the influence of different processing-energy factors at the rework steps (i.e., post-maturation and post-drying) under medium uncertainty (CV=0.3), reflecting the case company's current operational condition. For both approaches, the best-performing configurations are selected: optimal main-step processing-energy factors for *OPT* and optimal $\beta$ values for *SBA*, assuming an initial sensor distortion ($\alpha_0^v = 0.20$), to reflect standard sensor accuracy. The primary y-axis shows the minimum average cumulative energy input per batch, and the secondary y-axis displays the corresponding rework ratio.

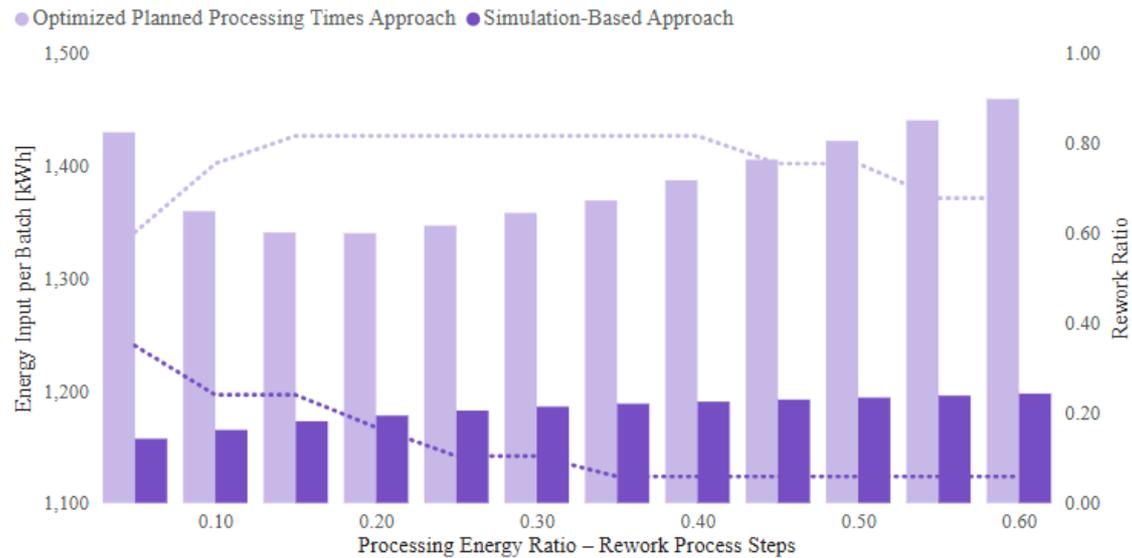

Figure 12: Impact of Sensor Distortion at Stopping Problem.

For *OPT*, the rework-step duration significantly affects energy consumption, with optimal configuration factor at approximately 0.2 of the expected energy. Conversely, *SBA* is largely insensitive to rework-step duration due to its flexibility in adjusting processing times via the energy safety parameter *β*. The best results are achieved when an optimal *β* is combined with a rework factor of approximately 0.05 of the expected energy. While the rework ratio remains consistently high for *OPT*, it decreases for *SBA* when additional energy is allocated to rework, highlighting the approach's flexibility and adaptability.

*7.5 Trade-Off Between Energy and Personnel Costs*

To extend our analysis, we convert energy consumption into monetary costs, combining them with labour costs for chamber inspections. Energy costs are calculated using Austria's average industrial electricity price of €0.197 per kWh for the second half of 2024, as reported by Eurostat. Personnel costs, also sourced from Eurostat, are €40.90 per hour, including direct and indirect expenses. Each rework inspection is estimated to take approximately 15 minutes, according to company experts.

*Figure 13* illustrates the Pareto front representing the trade-off between energy and personnel costs at the case company's current operational uncertainty (CV=0.3), assuming a standard initial sensor distortion ($\alpha_0^v = 0.20$) for the *SBA*. *OPT* configurations (i.e., solutions) appear in light purple, *SBA* configurations in dark purple, and Pareto-optimal solutions are marked with bigger data points. The overall optimum is highlighted, including total costs and the corresponding configuration (i.e., *β* and rework processing-energy factor).

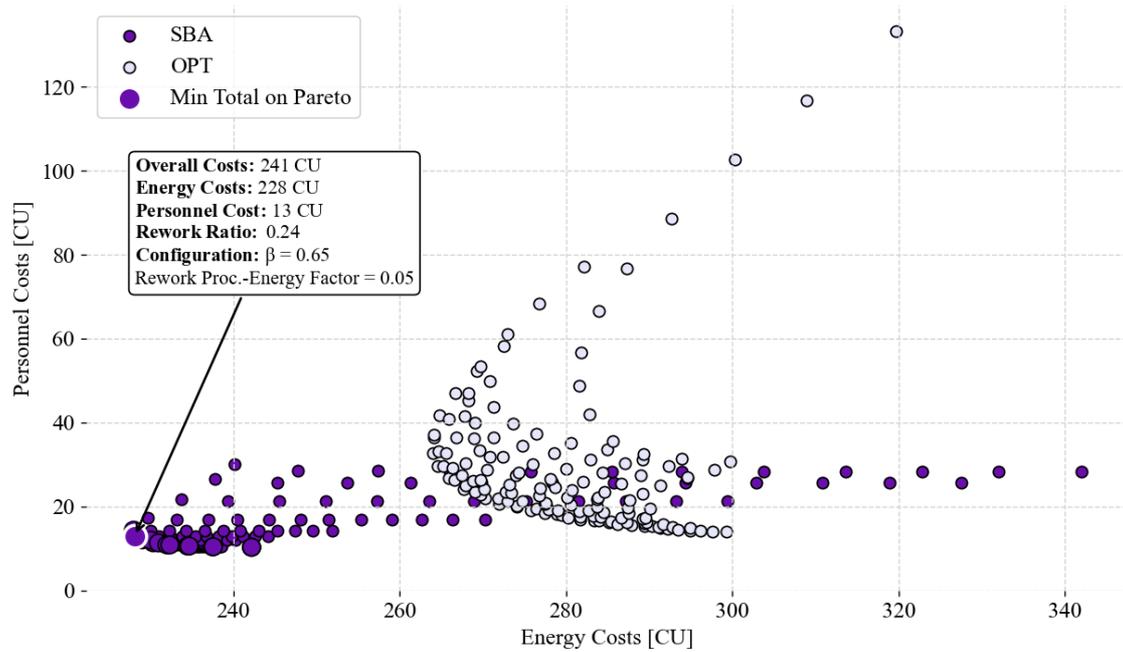

Figure 13: Pareto Front of Energy and Personnel Costs per Batch.

Heat-treatment planning has a substantial impact on overall operational costs, with poor parameter selection leading to significant financial consequences. The optimal configuration balances 13 CU in personnel costs – primarily from one mandatory inspection (see *Figure 2*) – with 228 CU in energy costs, yielding total batch costs of 241 CU. It corresponds to a moderately conservative safety margin ($β=0.65$), resulting in a rework ratio of about 0.24, and the shortest possible rework processing-energy factor of 0.05. Moreover, the analysis reveals that *SBA* consistently outperforms *OPT*, even under suboptimal configurations, highlighting its strong energy reduction potential. The results show that minimizing energy costs with *OPT* leads to significant additional personnel costs, which sheds light on the *Current Baseline Practise* of the case company, that accepts additional energy costs to reduce personnel costs. This trade-off is significantly improved with the developed *SBA* which underpins the energy reduction potential of applying a sensor based stopping policy in the case company. Therefore, the Pareto analysis provides actionable insights to the energy–labour cost trade-off, offering valuable decision support for industrial practitioners.

## 8. Conclusion

This paper addresses the urgent need for sustainable manufacturing practices by proposing a *Simulation-Based Approach (SBA)* for modelling stopping policies in stochastic production systems. Motivated by an energy-intensive heat-treatment problem at a lead-acid battery manufacturing operating under high uncertainty, the study demonstrates how sensor-driven process control and data-based decision-making can significantly reduce energy consumption.

The *SBA* iteratively adjusts batch-specific processing times based on simulated sensor data, estimating the energy requirement to meet product characteristics (i.e., stopping condition). A Markovian model captures realistic sensor behaviour and generates time-series data for Bayesian estimation of the energy requirement. These estimates define an evolving energy threshold, which is then used to adjust the remaining processing time, enabling precise and energy-efficient control actions.

Results show that *SBA* significantly reduces energy consumption compared to the *Optimized Planned Processing Times (OPT)*, which applies fixed planned processing times uniformly across all batches. At strict significance level $\alpha=0.01$, *SBA* consistently outperforms *OPT*. Relative to the case company's *Current Baseline Practice*, SBA reduces energy input by 14-25%, depending on process uncertainty and sensor accuracy. The approach maintains robust performance across varying environmental conditions by adjusting configuration parameters. In some cases, *SBA* even performs statistically equivalent to an *Ideal Scenario* with perfectly known energy requirements ($\alpha=0.01$). A Pareto analysis further highlights the value of *SBA* in balancing energy and inspection-labour costs, showing its superiority over *OPT* even under suboptimal configurations. Overall, the results demonstrate the *SBA's* effectiveness in a real industrial environment and its potential for broader application in sustainable sensor-driven production systems operating under uncertainty.

While developed for a specific industrial context, the *SBA* is designed to be generic and applicable to other stochastic production systems. Future research could expand the model to multi-chamber systems, where stopping decisions are integrated with batch scheduling, or apply the approach to other domains requiring real-time adaptive control under uncertainty. AI approaches such as reinforcement learning could also be investigated to derive the optimal stopping policy.

## Acknowledgments

This work was supported by the research project Sched-Energy [FO999891173] and the PhD-funding program EE-Scheduling [FO999905125]. The Sched-Energy project is funded by the Austrian Research Promotion Agency FFG in the program "Produktion der

Zukunft". EE-Scheduling is a PhD-funding program of the University of Applied Sciences Upper Austria, funded by the Austrian Research Promotion Agency FFG.

## Declaration of Interest Statement

The authors report there are no competing interests to declare.

## Data Availability Statement

The data that support the findings of this study are available in Zenodo at https://doi.org/10.5281/zenodo.15393963 upon reasonable request to the author Balwin Bokor.

## Appendix

Directly measuring the minimum energy required for curing and drying is impractical in the case company, so we estimated these values from routine metallic lead inspections carried out after maturation (*Figure 2*). During each inspection the proportion of free metallic lead in a plate is compared with a predefined threshold to assess chemical curing. Batches that fail are routed to post-maturation rework (*Figure 2*).

For every heat-treatment programme (i.e., Negative, Positive, Positive-with-Vaporisation, and Start/Stop), *Table 5* lists the company's current planned process times for maturation and drying, the corresponding cumulative energy input (derived from *Figure 3*), and the full set of inspection statistics. These are: sample size (number of inspections), mean, SD, CV of the metallic lead measurements, number of rework cases, and resulting rework ratio. No direct data exist for the Negative programme, so we assume it requires same or a slightly higher energy proportion as the Positive programme.

|  |  | Negative | Positive | Positive Vap. | Start/Stop | Wgt. Mean |
|---|---|---|---|---|---|---|
| Applied Planned Processing Time Maturation [h] |  | 20 | 36 | 36 | 24 |  |
| Cumulated Energy Input at Maturation [kWh] |  | 352 | 731 | 1,216 | 355 | 457 |
| Metallic Lead relative to Total Material [%] | Mean |  | 1.88 | 2.76 | 1.81 |  |
|  | SD |  | 0.61 | 0.66 | 0.52 |  |
|  | **CV** |  | **0.32** | **0.24** | **0.29** | **0.31** |
| Inspection Results | Nbr. Insp. |  | 183 | 145 | 115 |  |
|  | Requiring Rework |  | 61 | 15 | 8 |  |
|  | Rework Ratio |  | 0.33 | 0.10 | 0.07 |  |
| Estimations - Chemical Ripening | **Exp. Min. Energy [kWh]** | 308.84 | 641.36 | 934.08 | 249.14 | 380.50 |
|  | Higher Proportion Provided | 1.14 | 1.14 | 1.30 | 1.42 | 1.21 |
| Applied Planned Processing Time Drying [h] |  | 18 | 22 | 12 | 56 |  |
| Cumulated Energy Input at Drying [kWh] |  | 689 | 664 | 480 | 1,545 | 839 |
| Estimations - Humidity Level | **Exp. Min. Energy [kWh]** | 604.51 | 582.58 | 368.72 | 1,084.27 | 679.98 |
|  | Higher Proportion Provided | 1.14 | 1.14 | 1.30 | 1.42 | 1.21 |
| Expected Demand Share |  | 0.61 | 0.13 | 0.06 | 0.19 |  |

Table 5. Inspection Results for Heat-Treatment Programs.

We treat the metallic lead proportion as a proxy for the hidden minimum energy requirement. We further assume that fluctuations in required energy mirror those in the free metallic lead proportion and that the observed rework ratio corresponds to a specific percentile of this distribution. Under these assumptions, we back-calculate the expected minimum energy by evaluating the Gaussian inverse-CDF at the rework percentile, using the cumulative energy (accumulated values from *Figure 3*) provided during the planned processing time as the reference. The same logic, with identical ratios, is used to estimate the energy required to reach the target humidity level. *Table 2* reports the resulting programme-specific means, while the corresponding CVs are stress-tested in the numerical study. Multiplying each programme's frequency by its average lot size (both shown in *Table* 2) yields the demand shares in the bottom row of *Table 5*. Weighting the programme CVs with these shares yields an overall CV of ≈ 0.31, while the same weighting confirms that the firm's planned processing times supply, on average, 1.2 times the expected minimum energy. Together, these two weighted figures define the *Current Baseline Practice* benchmark used throughout the paper.